\documentclass[10pt,conference]{IEEEtran}
\pdfoutput=1
\usepackage[utf8]{inputenc}

\usepackage[hyphens]{url}
\usepackage[hidelinks]{hyperref}

\usepackage{complexity}

\usepackage{tikz-uml}

\usepackage{amsthm}
\theoremstyle{definition}
\newtheorem{exmp}{Example}[section]

\title{On Testing Quantum Programs}

\begin{document}
\bstctlcite{IEEEexample:BSTcontrol} 

\author{\IEEEauthorblockN{Andriy Miranskyy and Lei Zhang}
\IEEEauthorblockA{\textit{Department of Computer Science, Ryerson University, Toronto, Canada} \\
\{avm, leizhang\}@ryerson.ca}
}
\maketitle

\begin{abstract}
A quantum computer (QC) can solve many computational problems more efficiently than a classic one. The field of QCs is growing: companies (such as DWave, IBM, Google, and Microsoft) are building QC offerings. We position that software engineers should look into defining a set of software engineering practices that apply to QC's software. To start this process, we give examples of challenges associated with testing such software and sketch potential solutions to some of these challenges. 
\end{abstract}

\section{Introduction}\label{sec:intro}

A quantum computer (QC) can efficiently solve various problems that a classical computer (CC) cannot~\cite{feynman1982simulating}; this is known as the quantum supremacy~\cite{preskill2012quantum}. Examples of such problems (originating in various fields of science)  are scalable simulations of quantum systems in physics~\cite{feynman1982simulating}, efficient modelling of chemical reactions~\cite{aspuru2005simulated}, and fast breaking of encryption codes\footnote{ These encryption codes are based on integer factorization. Multiple QC-resistant encryption methods have been proposed~\cite{bernstein_post-quantum_2009}, but their implementation will require significant changes to various software: web browsers and web servers, mail and hard drive encryptors, etc. We conjecture that the amount of work necessary to introduce these changes into legacy software may be similar to that of the Y2K problem~\cite{britannica_y2k}.} in cryptography~\cite{shor1997}.

\subsubsection{QC underlying principles} A CC operates on a sequence of bits taking the values of $0$ or $1$.  A QC operated on quantum bits (hereon qubits), which are two-state quantum-mechanical systems~\cite{nielsen_chuang_2010}. Unlike the bit, the qubit is represented by a superposition of two states, inducing a probability distribution of the qubit's outcome being $0$ or $1$ upon measurement~\cite{nielsen_chuang_2010}. Essentially, the qubit is a stochastic system: every time we take a measurement on a given qubit, the probability of the outcome ($0$ or $1$) would change. The beauty of the qubits lies in the fact that qubits can be entangled, meaning that the number of states that a QC can represent is proportional to $2^q$, where $q$ is the number of qubits, whereas a CC operating on $q$ bits would be able to represent at most $q$ states~\cite{nielsen_chuang_2010}.

\subsubsection{QC applicability} Not every problem can benefit from the QC architecture: those problems falling under the bounded error quantum polynomial time ($\BQP$) class defined in computational complexity theory will~\cite{nielsen_chuang_2010}. The time complexity of algorithms, which solve $\BQP$ class problems, grows polynomially with the increase of the input size on a QC. On the contrary, the time complexity of the algorithms solving the same problems on a CC is not bounded above by a polynomial function and may grow faster (e.g., exponentially) with the increase of the size of the input. 

Formally, it was shown that the relations between $\BQP$ and other popular complexity classes are as follows: $\P \subseteq \BPP \subseteq \BQP \subseteq \P^{\#\P} \subseteq \PSPACE$, where $\P$ is a polynomial time complexity class, $\BPP$ is a bounded-error probabilistic polynomial time class, $\P^{\#\P}$ is $\P$ with $\#\P$ oracle class ($\#\P$ is a set of counting problems, and is a class of function problems rather than decision problems), and $\PSPACE$ is a polynomial space class, see~\cite{vazirani2002survey} for details. Currently, the consensus (although not formally proved) is that some of the nondeterministic polynomial time ($\NP$) problems do belong to the $\BQP$ set; however, $\BQP$ and $\NP$-complete sets of problems do not overlap (see~\cite{vazirani2002survey,nielsen_chuang_2010} for review), i.e., a QC will not be able to solve an $\NP$-complete problem.

\subsubsection{QC timeline} The field of quantum computing is young: Feynman introduced the idea of quantum computing in 1982~\cite{feynman1982simulating}; Shor proposed the first practically relevant algorithm (for breaking encryption protocols based on integer factorization) that can be efficiently computed on a QC in 1994~\cite{shor1997}.

A QC can be simulated on a CC~\cite{ibm_quantum,ms_quantum}. A quantum simulator interprets a mathematical function as part of a physical model~\cite{Johnson2014}; however, it will not yield performance improvement that a QC would provide, as the underlying host system of the simulator is still based on bits rather than qubits. Thus, one needs a real QC to reap performance benefits.

It took a while to implement an actual QC. A partnership between academia and IBM created the first working 2-qubit QC in 1998~\cite{Chuang1998}, but it took the company 18 years to make a 5-qubit QC accessible to the public in 2016~\cite{ibm2016}. 

At present, a few QCs are commercially available. DWave started selling adiabatic QC in 2011 (although debate about adiabatic QC being a ``true'' QC is ongoing\footnote{A hybrid of adiabatic and gate-based QC is promising~\cite{barends_digitized_2016}, but no commercial implementation is available.}~\cite{albash2017}) with the current offerings having $>$~2000 qubits. IBM gave access to  20- and 50-qubit gate-based superconducting QCs to academic and industrial partners to explore practical applications in 2017~\cite{ibm2017}. 

For non-commercial use, IBM offers 5- and 16-qubit QCs via IBM Q Experience online platform based on IBM Cloud (along with local- and Cloud-based simulators)~\cite{ibm_quantum}. Microsoft provides access to a simulator of a topological QC via Microsoft Quantum Development Kit~\cite{ms_quantum} (and is planning to give access to an actual QC in the future). Google built 72-qubit gate-based superconducting QC in 2018~\cite{google2018}, but it is not publicly accessible at the time of writing. 

\subsubsection{QC performance} When discussing the performance of the abovementioned QCs, we have to be mindful of the fact that the performance of the QCs (which are based on different architectures) cannot be compared merely based on the number of qubits that each QC has. Conceptually, it is similar to the fact that we cannot compare the performance of CC central processing unit (CPU) based solely on the number of CPU cores and the cores' frequency.  Standardization of benchmarks for QC is currently in the works by an IEEE Working Group~\cite{ieee_wg}. 

\subsubsection{QC adoption} Given that $\P \subseteq \BQP$~\cite{vazirani2002survey}, one may argue that QC will replace CC at some point in time. However, we conjecture that QC will not replace CC in the short run. Rather, QCs will be integrated into a System of Systems (SoS), where QC-based components will solve $\BQP$ problems (that CC cannot solve), while the solution will be passed to CC components for post-processing. Let us elaborate on this conjecture.

\begin{figure*}[t]
    \centering
    \resizebox{0.9\linewidth}{!}{
    \begin{tikzpicture}[outer/.style={draw=gray,dashed,fill=green!1,thick,inner sep=5pt}]
    \begin{umlseqdiag}
    \umlactor[scale = 0.5]{user}
    \umlbasicobject{WebApp's UI}
    \umlbasicobject{WebApp's Backend}
    \umlbasicobject{QC's Controller}
    \umlbasicobject{QC's Core}
    \begin{umlcall}[op=$N$, type=synchron, return=Return $\vec{L}$]{user}{WebApp's UI}
    \begin{umlcall}[op=$N$, type=synchron, return=Return $\vec{L}$]{WebApp's UI}{WebApp's Backend}
    \begin{umlcall}[op=QASM, type=synchron, return=Return $\vec{L}$]{WebApp's Backend}{QC's Controller}
    \begin{umlcall}[op=Control sequence, type=synchron, return= Return qubits' state]{QC's Controller}{QC's Core}
    \end{umlcall}
    \end{umlcall}
    \end{umlcall}
    \end{umlcall}
    \node (text1) [anchor=north] at ([xshift=5.5em, yshift=1.5em]WebApp's UI.north) {WebApp};
    \node (text2) [anchor=north] at ([xshift=5.5em, yshift=1.5em]QC's Controller.north) {QC};
    \begin{pgfonlayer}{background}
    \node[outer,fit=(WebApp's UI) (WebApp's Backend) (text1)] (A) {};
    \node[outer,fit=(QC's Controller) (QC's Core) (text2)] (B) {};
    \end{pgfonlayer}
    \end{umlseqdiag}
    \end{tikzpicture}
    }
    \caption{Sequence diagram for Example~\ref{ex:sos}. A user submits the value of integer $N$ for factorization via UI of a Web App, which passes $N$ to the WebApp's backend. At the backend where the value of $N$ is passed to, Shor’s algorithm is implemented using, say QisKit library~\cite{ibm_quantum}. The library translates QisKit code into Open Quantum Assembly Language (QASM)~\cite{cross2017open} and passes QASM listing to the Controller of a QC. The Controller, which initializes the QC Core based on the QASM code, triggers its execution and measures the values of the qubits once execution ends. The Controller converts the measurements into the elements of $\vec{L}$. These values are then returned to the Backend, UI, and, finally, the user. The sequence is depicted as synchronous, but can be made asynchronous if required by a use case. Note that the WebApp UI and Backend, as well as the QC Controller, are running on CCs. The QC Core represents the ``true'' QC. However, the QC Controller and the QC Core can be thought of as one QC system from practical perspective. }
    \label{fig:arch}
\end{figure*}
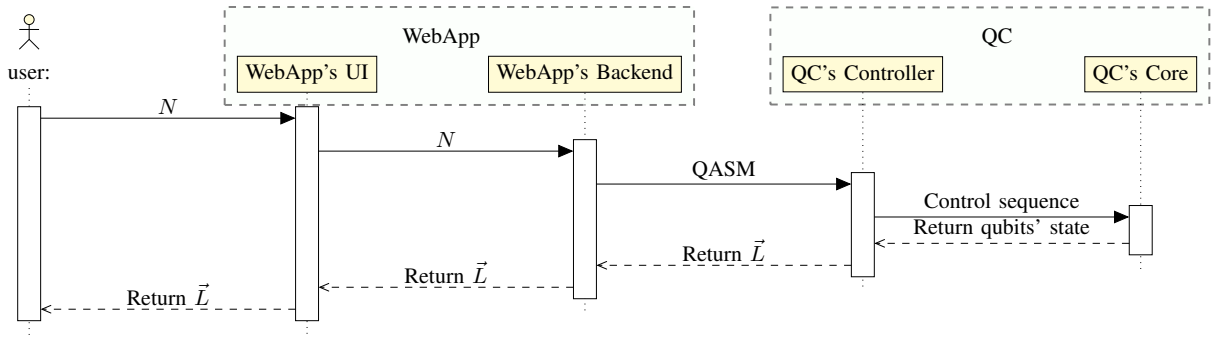

\begin{exmp}\label{ex:sos}
Suppose that we need to create a software-as-a-service for factoring large integers. The time complexity of the best algorithms available for a CC in the family of general number field sieves) is sub-exponential~\cite{pomerance96atale}. Thus, these algorithms will be ineffective for large integers. Instead, we will build a software component running Shor’s algorithm on a QC, which will be more efficient for large integers, because Shor’s algorithm computation time (as other $\BQP$ class algorithms) will grow polynomially with the growth of the input integer $N$ (when executed on a QC). The rest of the components, such as user interface (UI) and application program interface (API) for obtaining input (i.e., the value of $N$) to be passed to the QC component and to return the vector of factors $\vec{L}$ back to the user will be implemented on a CC, as depicted in Figure~\ref{fig:arch}. This is similar in nature to the existing Cloud solution for online access to IBM QCs~\cite{ibm_quantum}, where a Cloud-based interface for writing programs for a QC is running on a CC, while the program is then passed to the QC for execution.
\end{exmp}

Why cannot we implement all of this functionality on a QC? In the distant future, as equipment becomes cheaper to procure and operate, and the higher-level languages for QC are created, the replacement of CC with QC will become more probable. However, we hypothesize that it is not going to happen soon. Let us elaborate on the rationale for this statement. 
First, the QCs are expensive: e.g., DWave QC is valued at \$15 million~\cite{wired2017}. While other companies do not disclose their prices, we conjecture that the price tags of other QCs (magnitude-wise) are similar. Moreover, the operation of these computers requires cryogenic equipment (operated by highly qualified personnel), which further contributes to the costs. 

Second, modern QC programming languages, such as IBM QisKit Python package~\cite{ibm_quantum} and Microsoft Q\# language~\cite{svore2018}, operate at the level of qubits and quantum circuits. Creation of a UI and API in such a language would be very time-consuming and expensive. 

Notwithstanding, these languages integrate nicely into CC domain, simplifying the creation of SoS. IBM QisKit is implemented as a Python library, running on a CC. Once translated to QC machine language (via Open Quantum Assembly Language~\cite{cross2017open}), the code is passed to the QC for execution (the complexities of the call are encapsulated in the library’s code). Microsoft Q\# behaviour is similar: the code is developed on a CC and then passed to a QC for execution.

\subsubsection{Our position} Based on the above, QCs are becoming more mainstream; and we are not the only one making this claim, e.g., see~\cite{castelvecchi2017quantum}. Thus, we position that the Software Engineering (SE) community should start thinking about bringing SE practices into the domain of QCs. To do this, we need to answer some research questions. To name a few: which of the SE practices that we use in the CC domain can be ported to the QC domain; which of the practices are not applicable;  and which novel practices should be created to address QC domain challenges?

In this paper, to start the discussion, we focus on testing software created for QC. Due to a lack of space, we arbitrarily selected two topics related to testing activities: white- and black-box testing, discussed in Section~\ref{sec:box}, and verification and validation, discussed in Section~\ref{sec:vv}. Section~\ref{sec:conclusions} concludes the paper.

\section{White- and black-box testing} \label{sec:box}

Two widespread methods of testing are white-box and black-box testing. The former method tests internal data structures and program flow. The latter method tests the functionality, ignoring the inner workings of the software, answering the following question: will I get an expected output for a given input?

We can perform all of the standard white-box activities on the code listing, such as code reviews and code inspections. However, interactive debugging (another popular white-box activity) is challenging, because a QC is a black-box, by construction. Based on the classic quantum mechanics\footnote{Recently, Vijay et al.~\cite{vijay2012stabilizing} have invented a clever way of measuring a qubit state without ending superposition. It remains to be seen if this technology can be transferred to a QC. }, we cannot observe the inner working of a program (executed on a QC) without altering the program’s state and the final result, as measuring a qubit destroys superposition~\cite{vijay2012stabilizing}. 

This implies that, currently, we cannot perform interactive debugging of a program running on a QC, as we have to stop the program and take the measurements. Thus, we have to resort to black-box testing when dealing with a program running on an actual QC. Note that we can do white-box testing in QC simulators (e.g., we can use xUnit test framework to test Q\# programs~\cite{svore2018}).

\section{Verification and Validation}\label{sec:vv}

When testing the programs, how can we ensure that our code follows the design document and that the QC is doing what it is supposed to do? And even if our code reflects the design, how can we make certain that the output of the program delivers what a user needs? The former will be discussed in Section~\ref{sec:verification}, the latter --- in Section~\ref{sec:validation}.

\subsection{Verification}\label{sec:verification}

As discussed in Section~\ref{sec:box}, we can apply the full spectra of verification techniques on the code listings, but verification of a running program on an actual QC is more challenging, as we cannot use interactive debugger. To verify the correctness, we can try to run and debug our program in a local or online simulator, such as~\cite{ibm_quantum,ms_quantum}. However, as the simulators run on CC, we will have to limit the complexity of our test cases to obtain results within a reasonable amount of time. This will help us to eliminate some of the defects (a taxonomy of QC bugs is being developed~\cite{huang2018qdb}), but does not guarantee that no other defects will be encountered while solving production-scale problems. The same issue, conceptually, arises with CCs too, e.g., when dealing with buffer-overrun- and resource-leak-related defects.

The above test assumes that a simulator correctly and accurately resembles the actual QC, which is not always the case. Thus, a more definitive verification of correctness should be done on the the actual QC. Given the probabilistic nature of QC, we may have to execute the same code multiple times to increase the accuracy of our answer using Chernoff bound~\cite{nielsen_chuang_2010} (which is similar to probabilistic algorithms in $\BPP$ class running on CC~\cite{nielsen_chuang_2010}). This repeating functionality is built into packages like QisKit, but it requires a higher amount of computing resources (proportional to the number of repetitions).

The above approach assumes that the QC hardware, its operating system, and the compiler/translator of our program are running correctly, which is not always the case. To verify their correctness, we may need to execute the same program on multiple QCs (preferably from different manufacturers) and compare the results. If results differ significantly --- it may be a sign that there is an issue with one or more QCs under test. This is akin to correctness testing of a database engine by running the same query against different database engines~\cite{cialini2007method}.

A novel award-winning protocol, verifying QC computations with the help of a CC has been proposed~\cite{mahadev_2018}. It requires a significant amount of computational resources and, probably, will not be implemented shortly. However, as the computing power of QCs will increase, this protocol will become implementable in practice.

However, even if all of the above tests pass, it does not guarantee that the actual results returned by the QC are correct. This is where validation comes into play.

\subsection{Validation}\label{sec:validation}

When doing the validation, we need to make sure that the output of the algorithm satisfies the conditions provided in the requirements document (assuming that requirements were captured correctly). In other words, validation of a program running on a QC is similar to that of a program executed by a CC. Essentially, the ease of validation will depend on the difficulty of implementing a program for validating the results and the time needed\footnote{As discussed in Section~\ref{sec:intro}, many problems in $\BQP$ are solved efficiently on a QC, but are challenging to solve on a CC. However, the time needed to solve a problem is independent of the time needed to validate this solution.} to execute the validation. 

Before implementing a program, we need to estimate how long the validation process would take. To do so, we can resort to complexity analysis. Say, if the execution time of the validation\footnote{In the algorithm-related literature, the term verification rather than validation is used. We will use the term validation for consistency with the name of this section.} program would belong to $O(1)$, the validation process (given that it is easy to code up) would be straightforward. However, if the execution time would belong, say, to $O(n!)$, where $n$ would be proportionate to the length of input into the validation process (and to the length of the solution), then the validation process for a significantly large $n$ would be formidable. 

For simplicity, we can split the complexity of validation into two classes: polynomial time $\P$ bounded by $O(n^k)$ (i.e., validation time is bounded above asymptotically by $n^k$, where $k>0$) and super-polynomial time $\P^\C$, which is complementary to $\P$, bounded by $\omega(n^k)$ (i.e., validation time dominates asymptotically the $n^k$). We will look at examples of the algorithms belonging to these classes in subsections below.

Where should we implement the validation program: on a CC or a QC? In the program belongs to  $\P$ class, it can be implemented on either one, as $\P \subseteq \BQP$. However, as discussed in Section~\ref{sec:intro}, it is challenging to program a QC as we are dealing with low-level programming language. Moreover, the cost of running a QC in comparison with a CC is high. Thus, it is simpler and more economically feasible to implement a validation program on a CC, if possible.

In the case of $\P^\C$ class, the answer is less straightforward. If the size of the input into validation program is small, we may be able to still leverage a CC (especially if we can parallelize the validation on a CC cluster). However, we may have to resort to a QC for larger problems. If the validation program belongs to classes which are a subset of $\BQP$, such as $\BPP$ class, then QC is a good match. However, if the validation belongs to a ``harder’’ class, such as $\NP$-complete, then QC may also fail to deliver timely results. In this case, we may have to resort to a heuristic that tries to roughly validate the solution, but does not guarantee that solution is correct.

Let us look at one example of algorithms from both classes and ways to run the validation.

\subsubsection{Polynomial $\P$}

Shor’s integer factoring algorithm takes integer $N$ as input and returns a vector of prime factors $\vec{L}$ for $N$~\cite{shor1997}. The solution runs on a QC in polynomial time, $O((\log N)^2 (\log \log N) (\log\log\log N) )$ to be specific~\cite{shor1997}. The validation complexity is independent of the solution complexity,  growing linearly with the number of elements in $\vec{L}$, deemed $l$, as we simply need to multiply the elements in $\vec{L}$ to do the validation. That is, the complexity of validation of Shor’s algorithm is $O(l)$. Thus, we can easily\footnote{Although we may have to leverage existing libraries for multiplication of integers with arbitrary precision, such as Java BigInteger.} perform validation on a CC. Note that validation of Shor's algorithm can be carried on a QC, but this is economically inferior, as discussed above.

\subsubsection{Super-Polynomial $\P^\C$}

Boson sampling is a good example of a problem that is challenging to verify. Yet, the algorithm is crucial\footnote{It may lead to the implementation of a non-universal QC, which will still be more efficient than CC for some tasks, see~\cite{aaronson2011} for details.}. Experimentally, the algorithm is typically implemented using photons (belonging to the family of boson particles~\cite{nielsen_chuang_2010}). To implement the algorithm, we need a linear-optical circuit with $m$ modes that is injected with $h$ individual photons ($m>h$)~\cite{giordani_experimental_2018}. In this implementation, the boson sampling task reduces to creating a sample from the probability distribution of individual photon measurements at the circuit’s output. 

This algorithm cannot be computed on a CC for large values of $m$ and $h$, as it requires computing a permanent of a matrix which is a $\#\P$-hard problem~\cite{aaronson2011,valiant1979}. At best, it requires $O(h 2^h + mh^2)$ computations~\cite{clifford2018}.

However, the problem does fall~\cite{aaronson2011} into $\PostBQP$ class ($\BQP$ class with post-selection), which can be efficiently computed on a QC. Validation of the results on a CC is also a $\#\P$-hard problem, as we again need to compute the permanent of a matrix. However, one may adopt a heuristic to estimate goodness of findings (essentially, performing approximate validation) using machine learning approach~\cite{giordani_experimental_2018}.

In the above example, to perform an accurate validation, we need to do it on a QC. Ideally, this should be done on a different QC to simultaneously check the correctness of the computer itself (as was discussed in Section~\ref{sec:verification}). The code of the validation software would be similar to the one of the solution software. Thus, if resources permit, one may want to create the validation code from scratch (rather than reusing the existing code from the solution) to avoid migration of the defects from the solution code into validation code.

\section{Conclusions}\label{sec:conclusions}
QCs are becoming more mainstream. Thus, we position that software engineers should start the process of bringing SE practices into the domain of QCs. To do this, we need to identify which existing methods are transferable from the domain of CCs, which have to be altered, and which have to be created.

To start the discussion on this matter, we list a number of challenges associated with white- and black-box testing as well as verification and validation of programs running on a QC. We then list some of the existing SE practices that are readily transferable to the QC domain (e.g., code review),  some that are difficultly transferable (e.g., interactive debugging), and some that have to be introduced (e.g., complexity-dependent placement of a validation program).

We hope that this paper will catalyze the process of defining SE practices for QCs and that SE specialists in academia and industry will start exploring this fascinating area of computing, expanding our work to other areas of testing and the remaining phases of the software development life cycle.

\bibliographystyle{IEEEtran}
\bibliography{references} 

\begin{thebibliography}{10}
\providecommand{\url}[1]{#1}
\csname url@samestyle\endcsname
\providecommand{\newblock}{\relax}
\providecommand{\bibinfo}[2]{#2}
\providecommand{\BIBentrySTDinterwordspacing}{\spaceskip=0pt\relax}
\providecommand{\BIBentryALTinterwordstretchfactor}{4}
\providecommand{\BIBentryALTinterwordspacing}{\spaceskip=\fontdimen2\font plus
\BIBentryALTinterwordstretchfactor\fontdimen3\font minus
  \fontdimen4\font\relax}
\providecommand{\BIBforeignlanguage}[2]{{%
\expandafter\ifx\csname l@#1\endcsname\relax
\typeout{** WARNING: IEEEtran.bst: No hyphenation pattern has been}%
\typeout{** loaded for the language `#1'. Using the pattern for}%
\typeout{** the default language instead.}%
\else
\language=\csname l@#1\endcsname
\fi
#2}}
\providecommand{\BIBdecl}{\relax}
\BIBdecl

\bibitem{feynman1982simulating}
R.~P. Feynman, ``Simulating physics with computers,'' \emph{International
  journal of theoretical physics}, vol.~21, no. 6-7, pp. 467--488, 1982.

\bibitem{preskill2012quantum}
J.~Preskill, ``Quantum computing and the entanglement frontier,'' \emph{arXiv
  preprint arXiv:1203.5813}, 2012.

\bibitem{aspuru2005simulated}
A.~Aspuru-Guzik, A.~D. Dutoi \emph{et~al.}, ``Simulated quantum computation of
  molecular energies,'' \emph{Science}, vol. 309, no. 5741, pp. 1704--1707,
  2005.

\bibitem{bernstein_post-quantum_2009}
D.~J. Bernstein, J.~Buchmann, and E.~Dahmen, Eds., \emph{Post-{Quantum}
  {Cryptography}}.\hskip 1em plus 0.5em minus 0.4em\relax Berlin Heidelberg:
  Springer-Verlag, 2009.

\bibitem{britannica_y2k}
\BIBentryALTinterwordspacing
``{Y2K bug | Definition, Hysteria, \& Facts | Britannica.com},'' accessed on
  2018-09-25. [Online]. Available:
  \url{https://www.britannica.com/technology/Y2K-bug}
\BIBentrySTDinterwordspacing

\bibitem{shor1997}
\BIBentryALTinterwordspacing
P.~W. Shor, ``Polynomial-time algorithms for prime factorization and discrete
  logarithms on a quantum computer,'' \emph{SIAM J. Comput.}, vol.~26, no.~5,
  pp. 1484--1509, Oct. 1997. [Online]. Available:
  \url{http://dx.doi.org/10.1137/S0097539795293172}
\BIBentrySTDinterwordspacing

\bibitem{nielsen_chuang_2010}
M.~A. Nielsen and I.~L. Chuang, \emph{Quantum Computation and Quantum
  Information: 10th Anniversary Edition}.\hskip 1em plus 0.5em minus
  0.4em\relax Cambridge Univ. Press, 2010.

\bibitem{vazirani2002survey}
U.~Vazirani, ``A survey of quantum complexity theory,'' in \emph{Proceedings of
  Symposia in Applied Mathematics}, vol.~58, 2002, pp. 193--220.

\bibitem{ibm_quantum}
\BIBentryALTinterwordspacing
``{IBM Q Experience},'' accessed on 2018-09-25. [Online]. Available:
  \url{https://quantumexperience.ng.bluemix.net/qx/experience}
\BIBentrySTDinterwordspacing

\bibitem{ms_quantum}
\BIBentryALTinterwordspacing
``Quantum computing {\textbar} {Microsoft},'' accessed on 2018-09-25. [Online].
  Available: \url{https://www.microsoft.com/en-us/quantum/}
\BIBentrySTDinterwordspacing

\bibitem{Johnson2014}
T.~H. Johnson, S.~R. Clark, and D.~Jaksch, ``What is a quantum simulator?''
  \emph{EPJ Quantum Technology}, vol.~1, no.~1, p.~10, Jul 2014.

\bibitem{Chuang1998}
\BIBentryALTinterwordspacing
I.~L. Chuang, N.~Gershenfeld, and M.~Kubinec, ``Experimental implementation of
  fast quantum searching,'' \emph{Phys. Rev. Lett.}, vol.~80, pp. 3408--3411,
  Apr 1998. [Online]. Available:
  \url{https://link.aps.org/doi/10.1103/PhysRevLett.80.3408}
\BIBentrySTDinterwordspacing

\bibitem{ibm2016}
\BIBentryALTinterwordspacing
``{IBM News room - 2016-05-04 IBM Makes Quantum Computing Available on IBM
  Cloud to Accelerate Innovation - United States},'' accessed on 2018-09-25.
  [Online]. Available:
  \url{https://www-03.ibm.com/press/us/en/pressrelease/49661.wss}
\BIBentrySTDinterwordspacing

\bibitem{barends_digitized_2016}
\BIBentryALTinterwordspacing
R.~Barends, A.~Shabani \emph{et~al.}, ``Digitized adiabatic quantum computing
  with a superconducting circuit,'' \emph{Nature}, vol. 534, pp. 222--226, Jun.
  2016. [Online]. Available: \url{http://dx.doi.org/10.1038/nature17658}
\BIBentrySTDinterwordspacing

\bibitem{albash2017}
\BIBentryALTinterwordspacing
T.~Albash, V.~Martin-Mayor, and I.~Hen, ``Temperature scaling law for quantum
  annealing optimizers,'' \emph{Phys. Rev. Lett.}, vol. 119, pp.
  110\,502:1--110\,502:7, Sep 2017. [Online]. Available:
  \url{https://link.aps.org/doi/10.1103/PhysRevLett.119.110502}
\BIBentrySTDinterwordspacing

\bibitem{ibm2017}
\BIBentryALTinterwordspacing
``{IBM Announces Collaboration with Leading Fortune 500 Companies, Academic
  Institutions and National Research Labs to Accelerate Quantum Computing - Dec
  13, 2017},'' accessed on 2018-09-25. [Online]. Available:
  \url{https://newsroom.ibm.com/2017-12-13-IBM-Announces-Collaboration-with-Leading-Fortune-500-Companies-Academic-Institutions-and-National-Research-Labs-to-Accelerate-Quantum-Computing}
\BIBentrySTDinterwordspacing

\bibitem{google2018}
\BIBentryALTinterwordspacing
``{Google AI Blog: A Preview of Bristlecone, Google’s New Quantum
  Processor},'' accessed on 2018-09-25. [Online]. Available:
  \url{https://ai.googleblog.com/2018/03/a-preview-of-bristlecone-googles-new.html}
\BIBentrySTDinterwordspacing

\bibitem{ieee_wg}
\BIBentryALTinterwordspacing
``{P7131 - Standard for Quantum Computing Performance Metrics \& Performance
  Benchmarking},'' accessed on 2018-09-25. [Online]. Available:
  \url{https://standards.ieee.org/project/7131.html}
\BIBentrySTDinterwordspacing

\bibitem{pomerance96atale}
C.~Pomerance, ``A tale of two sieves,'' \emph{Notices Amer. Math. Soc},
  vol.~43, pp. 1473--1485, 1996.

\bibitem{cross2017open}
A.~W. Cross, L.~S. Bishop \emph{et~al.}, ``Open quantum assembly language,''
  \emph{arXiv preprint arXiv:1707.03429}, 2017.

\bibitem{wired2017}
\BIBentryALTinterwordspacing
``{D-Wave 2000Q goes on sale | WIRED UK},'' accessed on 2018-09-25. [Online].
  Available:
  \url{https://www.wired.co.uk/article/d-wave-2000q-quantum-computer}
\BIBentrySTDinterwordspacing

\bibitem{svore2018}
\BIBentryALTinterwordspacing
K.~Svore, A.~Geller \emph{et~al.}, ``{Q\#: Enabling Scalable Quantum Computing
  and Development with a High-level DSL},'' in \emph{Proceedings of the Real
  World Domain Specific Languages Workshop 2018}, ser. RWDSL2018.\hskip 1em
  plus 0.5em minus 0.4em\relax New York, NY, USA: ACM, 2018, pp. 7:1--7:10.
  [Online]. Available: \url{http://doi.acm.org/10.1145/3183895.3183901}
\BIBentrySTDinterwordspacing

\bibitem{castelvecchi2017quantum}
D.~Castelvecchi, ``Quantum computers ready to leap out of the lab in 2017,''
  \emph{Nature News}, vol. 541, no. 7635, pp. 9--10, 2017.

\bibitem{vijay2012stabilizing}
R.~Vijay, C.~Macklin \emph{et~al.}, ``{Stabilizing Rabi oscillations in a
  superconducting qubit using quantum feedback},'' \emph{Nature}, vol. 490, pp.
  77--80, 2012.

\bibitem{huang2018qdb}
Y.~Huang and M.~Martonosi, ``QDB: from quantum algorithms towards correct
  quantum programs,'' \emph{arXiv preprint arXiv:1811.05447}, 2018.

\bibitem{cialini2007method}
E.~Cialini, A.~Loreto, and D.~Godwin, ``Method, system, and program for
  determining discrepancies between database management systems,'' 2007, {US
  Patent App. US20070100783A1}.

\bibitem{mahadev_2018}
U.~Mahadev, ``Classical verification of quantum computations,'' in \emph{2018
  IEEE 59th Annual Symposium on Foundations of Computer Science (FOCS)}, Oct
  2018, pp. 259--267.

\bibitem{aaronson2011}
\BIBentryALTinterwordspacing
S.~Aaronson and A.~Arkhipov, ``The computational complexity of linear optics,''
  in \emph{Proceedings of the Forty-third Annual ACM Symposium on Theory of
  Computing}, ser. STOC '11.\hskip 1em plus 0.5em minus 0.4em\relax ACM, 2011,
  pp. 333--342. [Online]. Available:
  \url{http://doi.acm.org/10.1145/1993636.1993682}
\BIBentrySTDinterwordspacing

\bibitem{giordani_experimental_2018}
\BIBentryALTinterwordspacing
T.~Giordani, F.~Flamini \emph{et~al.}, ``Experimental statistical signature of
  many-body quantum interference,'' \emph{Nature Photonics}, vol.~12, no.~3,
  pp. 173--178, Mar. 2018. [Online]. Available:
  \url{https://doi.org/10.1038/s41566-018-0097-4}
\BIBentrySTDinterwordspacing

\bibitem{valiant1979}
\BIBentryALTinterwordspacing
L.~Valiant, ``The complexity of computing the permanent,'' \emph{Theoretical
  Computer Science}, vol.~8, no.~2, pp. 189 -- 201, 1979. [Online]. Available:
  \url{http://www.sciencedirect.com/science/article/pii/0304397579900446}
\BIBentrySTDinterwordspacing

\bibitem{clifford2018}
\BIBentryALTinterwordspacing
P.~Clifford and R.~Clifford, ``The classical complexity of boson sampling,'' in
  \emph{Proceedings of the 29th Annual ACM-SIAM Symposium on Discrete
  Algorithms}, ser. SODA '18, 2018, pp. 146--155. [Online]. Available:
  \url{http://dl.acm.org/citation.cfm?id=3174304.3175276}
\BIBentrySTDinterwordspacing

\end{thebibliography}

\end{document}